\documentclass[aps,pra,twocolumn,groupedaddress,showpacs,amsfonts,amssymb]{revtex4}

\usepackage{graphics}
\usepackage{epsfig}
\usepackage{amsmath}
\usepackage{amsthm}

\begin{document}

\title{Stochastic resonance effects in quantum channels}
\author{Garry Bowen}
\email{gab30@damtp.cam.ac.uk}
\affiliation{Centre for Quantum Computation, DAMTP, 
University of Cambridge, Cambridge CB3 0WA, UK}
\author{Stefano Mancini}
\email{stefano.mancini@unicam.it}
\affiliation{Dipartimento di Fisica, Universit\`{a} di Camerino, 
I-62032 Camerino, Italy}

\begin{abstract}
We provide some examples of quantum channels where 
the addition of noise is able to enhance the information transmission rate.
This may happen for both quantum and classical uses and
realizes stochastic resonance effects.
\end{abstract}


\maketitle

\section{introduction}
Stochastic resonance is a phenomenon concerned about amplifying a small 
signal forcing a nonlinear system by addition of a stochastic
noise to the signal \cite{GHJM}. 
It may represent an example against common wisdom where the noise becomes fruitful for information transmission.
When considering aperiodical signals for a channel performance, 
the peak of mutual information between the input signal and output
signal is used as the definition of resonance \cite{BZ}.

Stochastic resonance has been already studied for some quantum systems \cite{GHJM}.
Nevertheless, due to the increasing attention dedicated to 
information transmission through quantum channels \cite{Keyl},
it becomes interesting to see whether such
effect also exits there.
The possibility was put forward in Ref.\cite{Ting} without however reaching any evidence, whilst Ref.\cite{PLA} showed a small enhancement of the classical information transmission rate by the addition of noise  in presence of threshold detection.

Here we clearly show stochastic resonance effects in quantum channels.
To this end we shall provide examples where 
the addition of noise is able to enhance the rate of information transmission.
This may happen for both quantum and classical uses of the channels.

\section{the Noisy Channel Model}

A quantum channel  is any completely positive and trace preserving (CPTP) map
from a set of input density operators $Q$ to a output one $Q'$ \cite{Keyl}.  
It can be described by a 
quantum operation ${\cal N}$ transforming
\begin{eqnarray} 
\rho_Q \stackrel{{\cal N}}{\longrightarrow} \rho_{Q'}\equiv{\cal N}(\rho_Q).
\label{CPTP}
\end{eqnarray}
One can represents the map ${\cal N}$ through a
unitary transformation
on a larger quantum system that includes
the environment $E$ of the system,
\begin{equation}
        {\cal N} (\rho_{Q}) = {\rm Tr}_{E}\left[ U_{QE} \left (
                                  \rho_{Q} \otimes |0\rangle_{E}\langle 0| \right )
                                  {U_{QE}}^{\dagger}\right] .
\label{Urep}
\end{equation}
where the environment has been considered
initially in a pure state $|0\rangle_{E}$ without
loss of generality, and the trace ${\rm Tr}_E$ is taken over environmental degree of freedom.

Alternatively, one can represents the map ${\cal N}$ through the
Kraus decomposition \footnote{This terminology arose because of the Kraus' book \cite{Kraus} where the decomposition appeared, however it was first proposed in Ref.\cite{Sud}.}
\begin{equation}
{\cal N} (\rho_Q)=\sum_i A_i\rho_Q A_i^\dagger\;,
\label{KSrep}
\end{equation}
where the operators $A_i$ satisfy the completeness relation
$\sum_i A_i^\dagger A_i = \mathbb{I}$.

For a classical use of the channel the transmission rate is determined by the Holevo-$\chi$ quantity \cite{Keyl}
\begin{equation}\label{chi}
\chi\equiv S\left({\cal N}(\rho_Q)\right)-\sum_i w_i S\left({\cal N}(\rho_Q^i)\right)\,,
\end{equation}
where $\sum_i w_i\rho_Q^i$ is a covex decomposition of $\rho_Q$ over the states $\rho_Q^i$ encoding the classical alphabet symbols. $S$ denotes the
von Neumann entropy.

The input-output mutual information reads
\begin{eqnarray}
S(Q:Q') = S(\rho_Q)+S({\cal N}(\rho_Q))-S(\rho_Q,{\cal N}),
\label{qmutual}
\end{eqnarray}
where 
\begin{eqnarray}
S(\rho_Q,{\cal N}) \equiv S(W)\equiv - {\rm Tr} (W \log_2 W), 
\label{Senv}
\end{eqnarray}
with 
\begin{eqnarray} 
W_{ij} \equiv \mbox{Tr}(A_i \rho_Q A_j^{\dagger}),
\label{W}
\end{eqnarray}
measures the amount of information 
exchanged between the system $Q$ and the environment $E$ during
their interaction \cite{Keyl}. 
This can be used to characterize the amount of quantum noise in the channel.
If the environment is initially in a pure state,
the entropy exchange is just the environment's entropy after the
interaction.

For a quantum use of the channel the transmission rate is determined by the 
 coherent information \cite{Keyl}
\begin{eqnarray}
I(\rho_Q,{\cal N}) \equiv S \left(
        {\cal N}(\rho_Q) \right) -
        S (\rho_Q,{\cal N}),
\label{cohinf}
\end{eqnarray}
which plays a role in quantum information theory analogous to that played
by the mutual information in classical information theory.
It can be positive, negative, or zero.

\section{The Amplitude Damping Channel}

Let us consider a modified amplitude damping channel as follows:
\begin{eqnarray}
A_1&=&\frac{\mathbb{I}+Z}{2}+\sqrt{1-p}\,\frac{\mathbb{I}-Z}{2}\;,
\label{ad1}\\
A_2&=&
\sqrt{p(1-q)}\,\frac{X+iY}{2}\;,
\label{ad2}\\
A_3&=&\sqrt{ pq}\,\frac{\mathbb{I}-Z}{2}\;,
\label{ad3}
\end{eqnarray}
where $\mathbb{I}$ is the identity matrix and $X$, $Y$,
$Z$ are the usual Pauli operators.
This channel has a simple interpretation:  when $q=0$ it behaves like an amplitude damping channel;
when $q\neq 0$ it also introduces a phase flip, and for $q=1$ it becomes a phase flip channel \cite{Keyl}. 
There are already some indications that these two kind of noise can compete each other \cite{MB}, 
thus improving the performance of the channel.

A general (input) state in the Bloch sphere representation
can be written as
\begin{equation}
\rho_Q=\frac{1}{2}\Big(\mathbb{I} + \vec{a}\cdot\vec{\sigma}\Big)\;.
\label{input}
\end{equation}
Here, $\vec{a}=(a_1,a_2,a_3)$ is the Bloch vector of length unity or less,
and $\vec{\sigma}$ is the vector of Pauli matrices. The length $\|\vec{a}\|$ of the vector $\vec{a}$
describes the purity of the input state.
The action of the channel on this state is:
\begin{equation}
{\cal N}(\rho_Q)=
\frac{1}{2}\Big(\mathbb{I} + \vec{b}\cdot\vec{\sigma}\Big)\;,
\end{equation}
in which 
\begin{equation}
\vec{b}=\Big(a_1 \sqrt{1-p},\,a_2 \sqrt{1-p},\, p-pq+a_3 (1-p+pq)\Big)\;.
\end{equation}
The matrix $W$ of Eq.(\ref{W}) thus computed reads
\begin{eqnarray}
W_{11}&=&\frac{1}{2}\left(2-p+a_3 p\right) \\
W_{12}&=&\frac{1}{2}\left(a_1-ia_2\right) \sqrt{p (1+p)(1-q)} =W_{21}^*\\
W_{13}&=&\frac{1}{2}\left(1- a_3\right) \sqrt{pq (1-p)}=W_{31}^*\\
W_{22}&=&\frac{1}{2}\left(1-a_3\right) p(1-q)\\ 
W_{23}&=& 0 =W_{32}^* \\
W_{33}&=&\frac{1}{2}\left(1-a_3\right)pq 
\end{eqnarray}
The noise strength becomes
$S(W)  = - \sum_{i=1}^3 \lambda_i \log_2 \lambda_i$,
with $\lambda_i$ the eigenvalues of the $W$ matrix. Furthermore, the eigenvalues of 
${\cal N}(\rho_Q)$ result 
\begin{equation}
\theta_{1,2}= \frac{1}{2}\left(1 \pm \left\| \,\vec{b} \, \right\| \right)\,,
\end{equation}
 then $S ({\cal N}(\rho_Q)) = - \sum_{i=1}^2 \theta_i \log_2 \theta_i$.

\begin{figure}
\epsfxsize=6.5cm\epsfbox{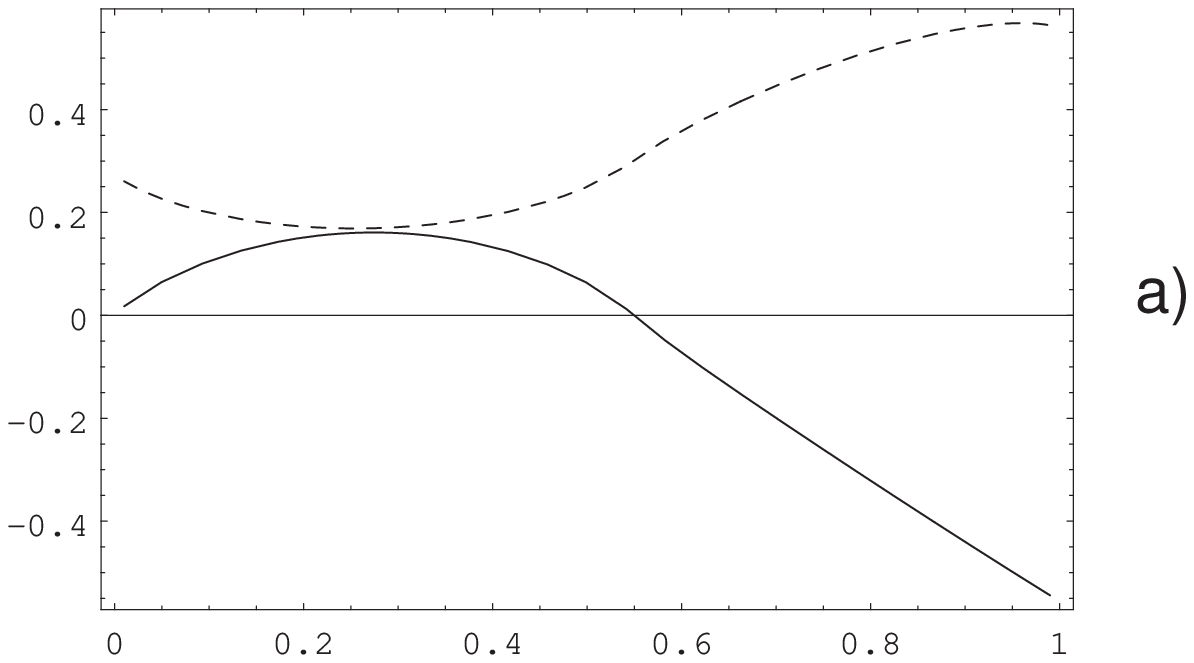}
\vskip 0.2cm
\epsfxsize=6.5cm\epsfbox{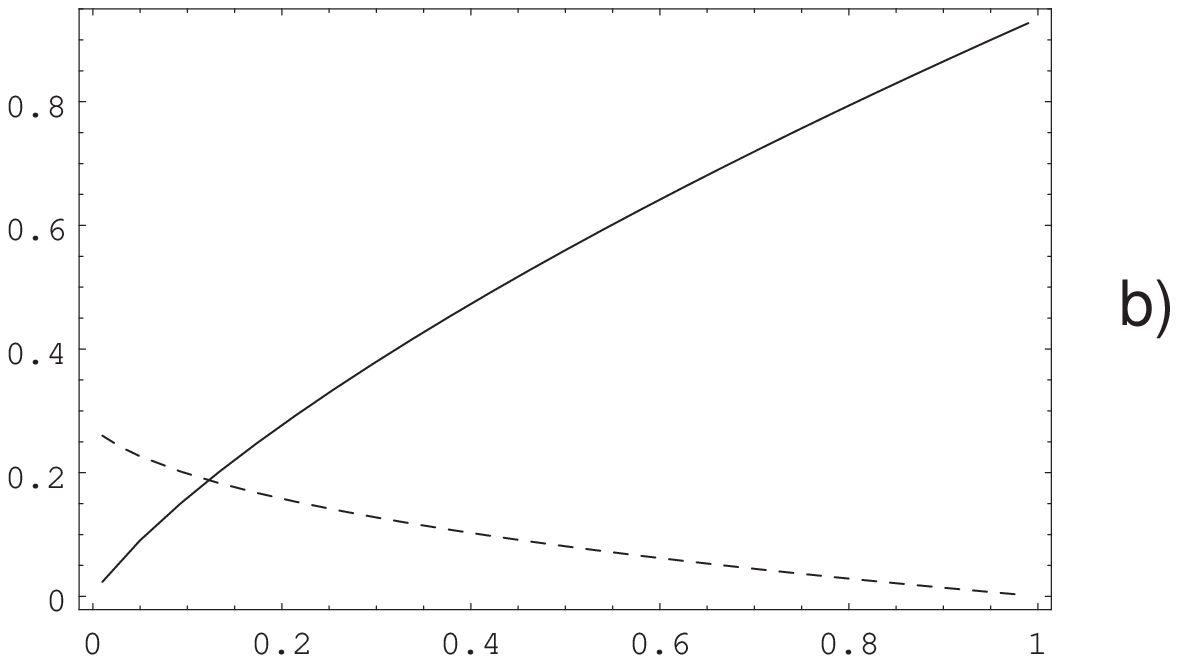}
\vskip 0.2cm
\epsfxsize=6.5cm\epsfbox{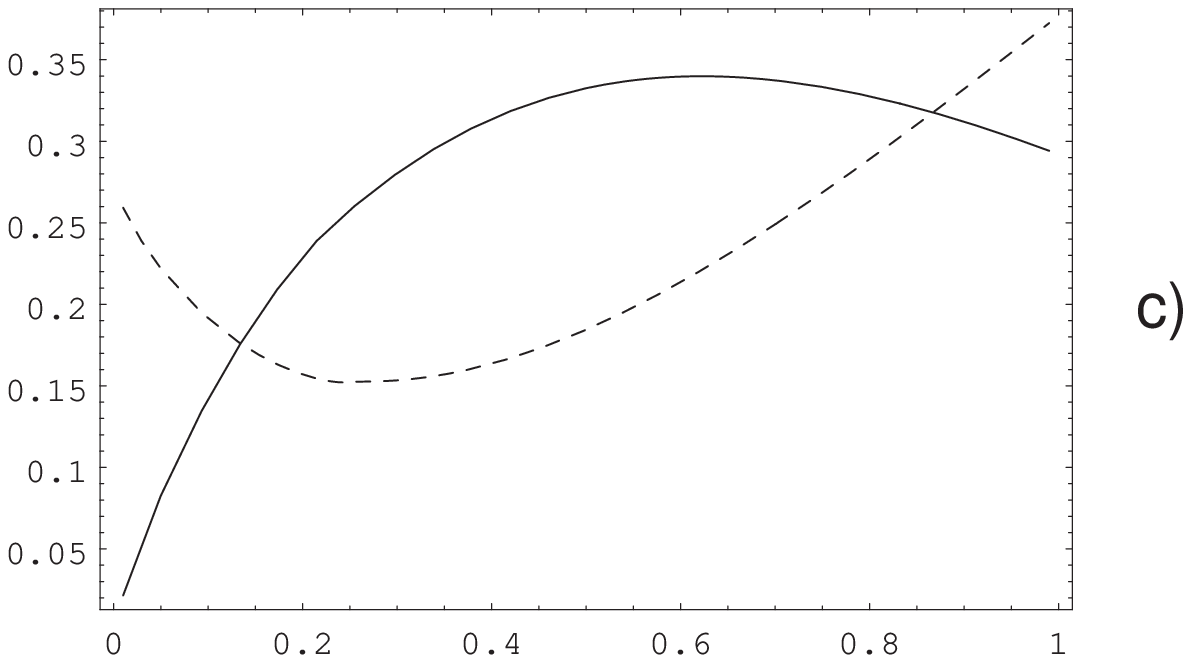}
\vskip 0.3cm
\caption{Plots of coherent information $I$ (solid lines) and noise $S(W)$ (dashed lines)
versus $p$ for $q=0$ (a), $q=1$ (b) and $q=0.5$ (c).
The input state is taken with:
$a_1 = 0.8$, $a_2 = 0.3$, $a_3 =0.3$, corresponding to a purity of $0.91$. 
}
\label{fig1}
\end{figure}

The noise enhancement can be investigated by looking for
some initial states, $(a_1, a_2, a_3)$, and damping rate, $p$,
where the slope of ${\partial I}/{\partial S(W)} >   0$.
However, in practice it is difficult to calculate such function analytically
and obtain useful results.
Some examples are given in Fig.~\ref{fig1}.
In particular, Fig.~\ref{fig1}a), Fig.~\ref{fig1}b) refer respectively to the limit cases of amplitude damping and phase flip channel where coherent information always decreases by increasing the noise. 
Instead, Fig.~\ref{fig1}c) referes to a mixing of the two channels and one can clearly see an enhancement of coherent information for increasing noise (e.g. for $0.25<p<0.60$).
It is worth noting that the stochastic resonance effect manifests only for high enough purity of the input states.

\section{The Bit Flip Channel}

Let us now consider a modified bit flip channel as follows:
\begin{equation}
A_1=\sqrt{1-p}\,\mathbb{I}\;,\;
A_2=
\sqrt{p(1-q)}\,X\;,\;
A_3=\sqrt{ pq}\,Z\;.
\end{equation}
Also this channel has a simple interpretation:  when $q=0$ it behaves like a bit flip channel;
when $q\neq 0$ it also introduces a phase flip, and for $q=1$ it becomes a phase flip channel. 

\begin{figure}
\epsfxsize=6.5cm\epsfbox{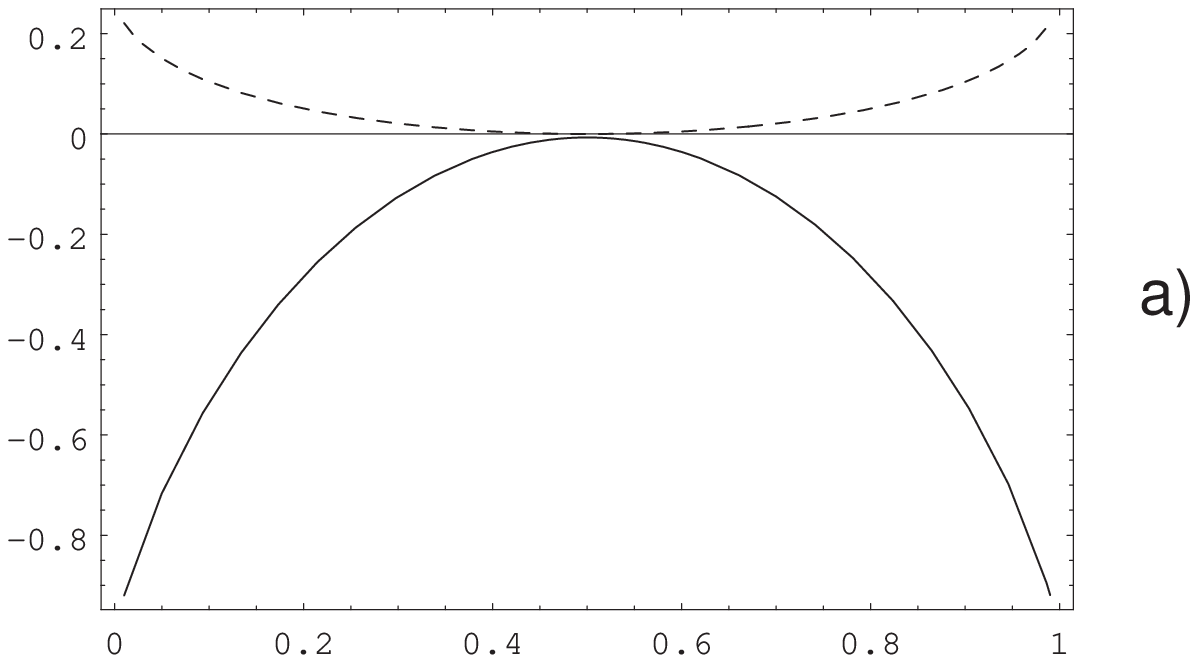}
\vskip 0.2cm
\epsfxsize=6.5cm\epsfbox{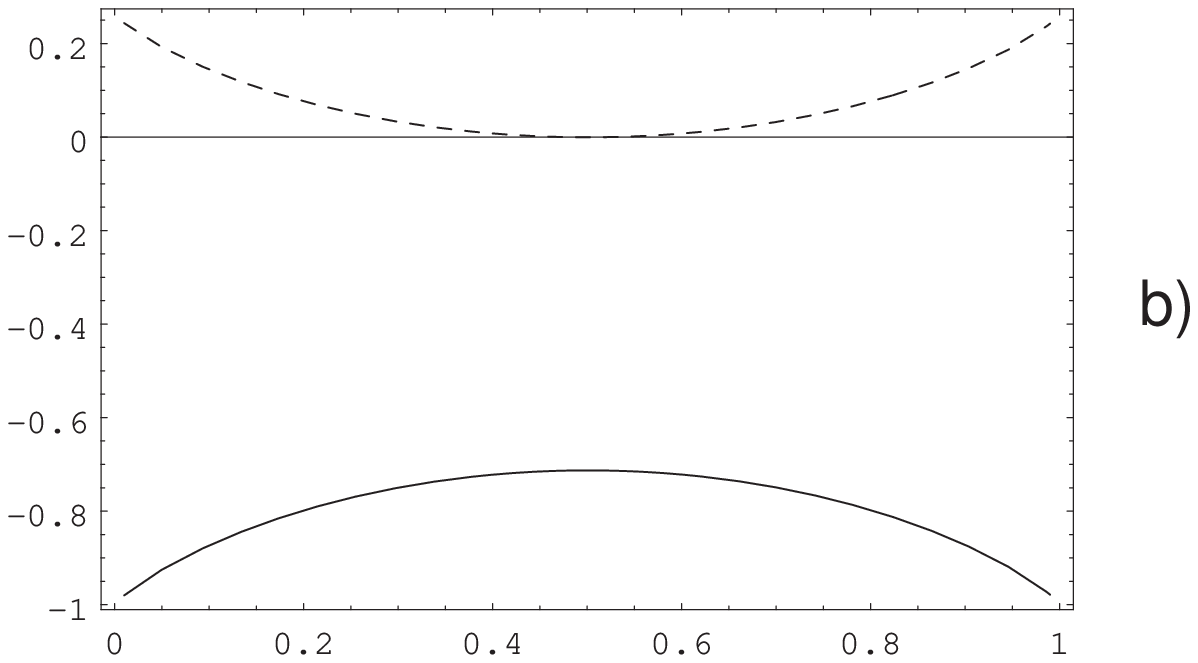}
\vskip 0.2cm
\epsfxsize=6.5cm\epsfbox{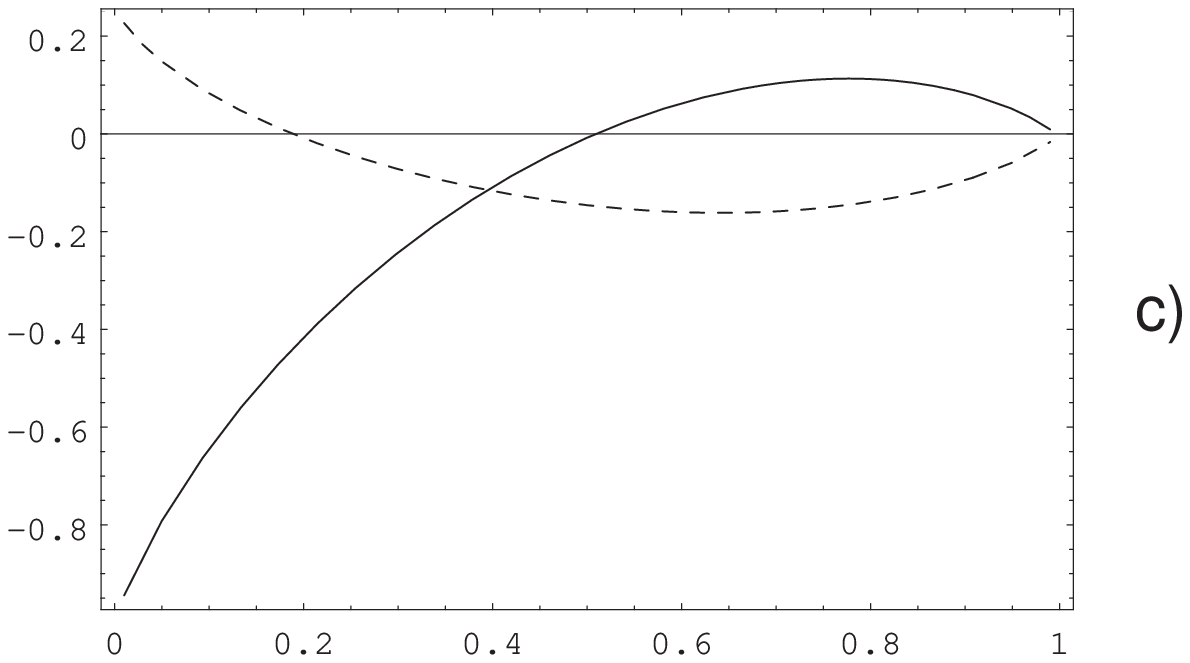}
\vskip 0.3cm
\caption{Plots of coherent information $I$ (solid lines) and noise $S(W)-1$ (dashed lines)
versus $p$ for $q=0$ (a), $q=1$ (b) and $q=0.5$ (c).
The input state is taken with:
$a_1 = 0.1$, $a_2 = 0.1$, $a_3 =0.9$, corresponding to a purity of $0.91$. 
}
\label{fig2}
\end{figure}

The action of the channel on the input state (\ref{input}) is:
\begin{equation}
{\cal N}(\rho_Q)=
\frac{1}{2}\Big(\mathbb{I} + \vec{b}\cdot\vec{\sigma}\Big)\;,
\end{equation}
in which 
\begin{equation}
\vec{b}=\Big(a_1 (1-2pq),\,a_2 (1-2p),\, a_3 (1-2p+2pq)\Big)\;.
\end{equation}
The matrix $W$ of Eq.(\ref{W}) thus computed reads
\begin{eqnarray}
W_{11}&=&1-p \\
W_{12}&=& a_1 \sqrt{p (1-p)(1-q)} =W_{21}^*\\
W_{13}&=& a_3 \sqrt{ pq (1-p)}=W_{31}^*\\
W_{22}&=& p(1-q)\\ 
W_{23}&=& ia_2\sqrt{p^2q(1-q)}=W_{32}^* \\
W_{33}&=& pq 
\end{eqnarray}
Then, by repeating the steps of the previous Section we can look for the desired effect.
Some examples are given in Fig.~\ref{fig2}.
In particular, Fig.~\ref{fig2}a), Fig.~\ref{fig2}b) refer respectively to the limit cases of bit flip and phase filip channel where the concavity of the coherent information is exactly opposite to that of the noise. 
Instead, Fig.~\ref{fig2}c) referes to a mixing of the two channels and one can clearly see an enhancement of coherent information for increasing noise (e.g. for $0.64<p<0.84$).
Also in this example the stochastic resonance effect manifests only for high enough purity of the input states.

\section{The Dense Coding Protocol}

Let us finally consider a modified dense coding protocol with the addition of some noise in the noiseless  channel used by Alice to send her qubit to Bob.

Let $\rho_{AB}$ be the state shared a priori by Alice and Bob, we consider 
as input state $\rho_Q=\sum_{i=1}^4w_i U_i\rho_{AB}U_i^{\dag}$, with $w_i=1/4$ and $U_1=\mathbb{I}_A$,  $U_2=X_A$,  $U_2=Y_A$, $U_2=Z_A$ like in the standard dense coding protocol \cite{garry}.
Then the transmission rate, according to Eq.(\ref{chi}), results \cite{garry}
\begin{equation}\label{Cdc}
\chi=1+S(\rho_B)-S\left(\left({\cal N}_A\otimes \mathbb{I}_B\right)\rho_{AB}\right)\,,
\end{equation}
where $\rho_B\equiv{\rm Tr}_A(\rho_{AB})$ and  ${\cal N}_A$
is the local map describing the noise according to Eq.(\ref{CPTP}).

Suppose now that $\rho_{AB}$ is a Werner-like mixture \cite{Wer}
\begin{equation}
\rho_{AB}=\frac{q}{4}\mathbb{I}_{AB}+(1-q)|\Psi^+\rangle_{AB}\langle\Psi^+|\,,
\end{equation}
where 
$|\Psi^+\rangle_{AB}\equiv\left(|0\rangle_A|0\rangle_B+|1\rangle_A|1\rangle_B\right)/\sqrt{2}$
is one of the Bell states and $q$ the degree of mixedness.
In such a case $\rho_B=\mathbb{I}_B/2$ hence $S(\rho_B)=1$.
Consider the noise ${\cal N}_A$ specifically described by operators
\begin{eqnarray}
A_1&=&\frac{\mathbb{I}+Z}{2}+\sqrt{1-p}\,\frac{\mathbb{I}-Z}{2}\;,\\
A_2&=&
\sqrt{p}\,\frac{X+iY}{2}\,,
\end{eqnarray}
on Alice's qubit. By comparison with Eqs.(\ref{ad1})-(\ref{ad2}) the above equations represent an amplitude damping channel.
Then, simple calculations give
\begin{eqnarray}
&&\left({\cal N}_A\otimes \mathbb{I}_B\right)\rho_{AB}=\frac{2-q(1-p)}{4}
|0\rangle_A\langle 0|  \otimes  |0\rangle_B\langle 0|\nonumber\\
&&+\frac{2p+q(1-p)}{4}|0\rangle_A\langle 0|   \otimes  |1\rangle_B\langle 1|\nonumber\\
&&+\frac{q(1-p)}{4}|1\rangle_A\langle 1|  \otimes  |0\rangle_B\langle 0|\nonumber\\
&&+\frac{(2-q)(1-p)}{4}|1\rangle_A\langle 1|   \otimes  |1\rangle_B\langle 1|\nonumber\\
&&+\frac{2(1-q)\sqrt{1-p}}{4}|0\rangle_A\langle 1|  \otimes  |0\rangle_B\langle 1|
\nonumber\\
&&+\frac{2(1-q)\sqrt{1-p}}{4}|1\rangle_A\langle 0|  \otimes  |1\rangle_B\langle 0|\,,
\end{eqnarray}
whose eigenvalues are
\begin{eqnarray}
\xi_{1,2}&=&\frac{1}{4}\left[q+p(1-q\mp 1)\right]\,,\\
\xi_{3,4}&=&\frac{1}{4}\left[2\mp\sqrt{p^2+4(1-p)(1-q)^3}-p(1-q)-q\right]\,.
\nonumber
\end{eqnarray}
Thus $S\left(\left({\cal N}_A\otimes \mathbb{I}_B\right)\rho_{AB}\right)=-\sum_{i=1}^4\xi_i\log\xi_i$.

\begin{figure}
\epsfxsize=6.5cm\epsfbox{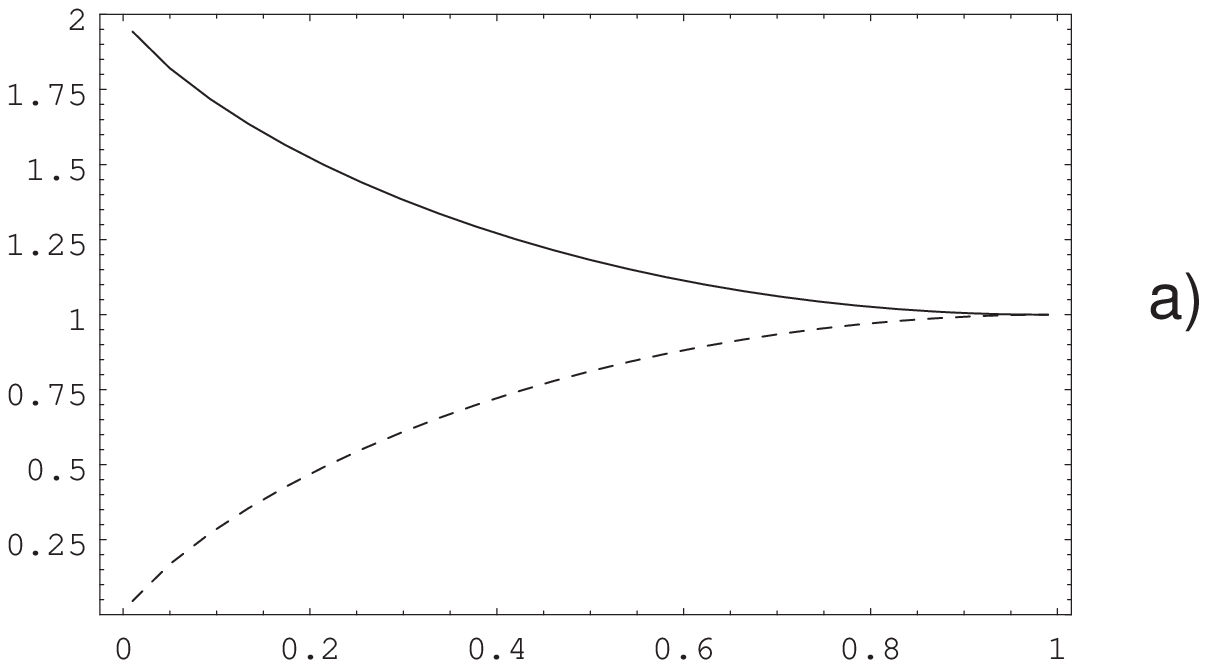}
\vskip 0.2cm
\epsfxsize=6.5cm\epsfbox{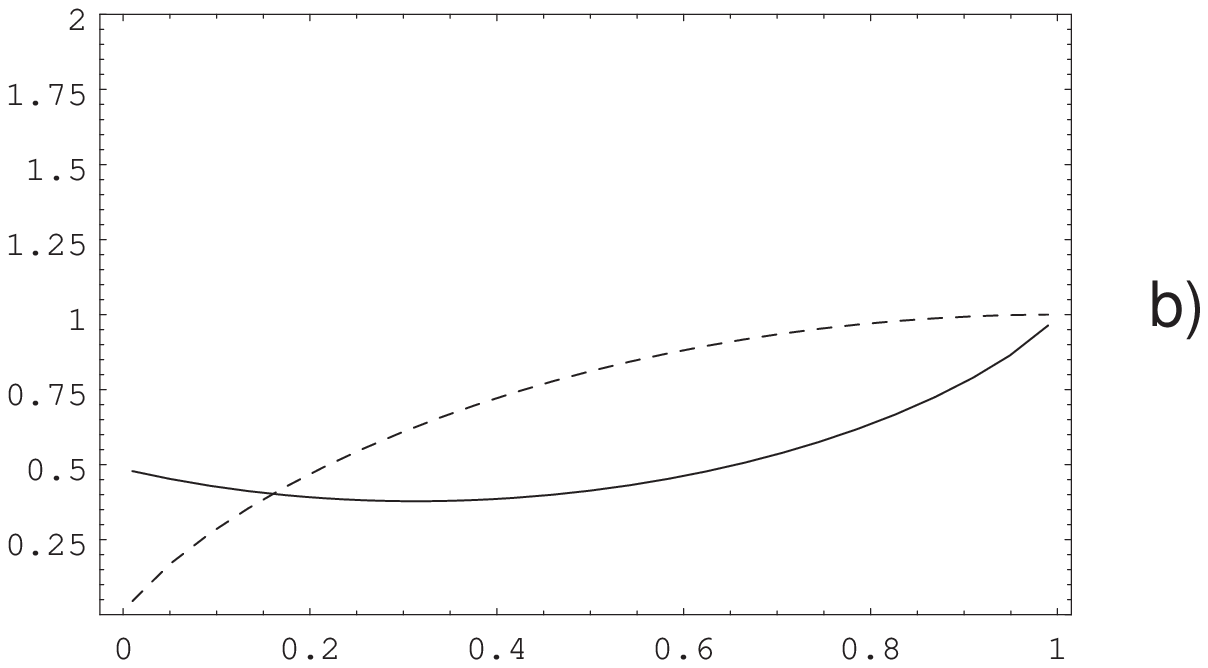}
\vskip 0.2cm
\epsfxsize=6.5cm\epsfbox{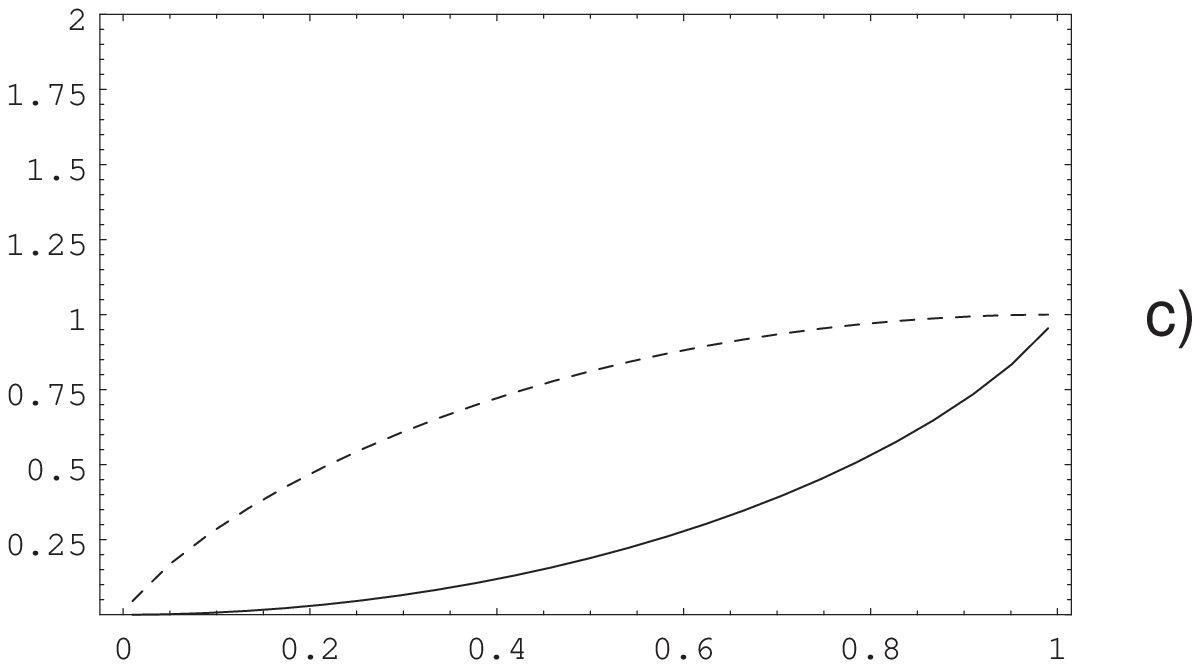}
\vskip 0.3cm
\caption{Plots of $\chi$ (solid lines) and noise $S(W)$ (dashed lines)
versus $p$ for $q=0$ (a), $q=0.5$ (b) and $q=1$ (c).
}
\label{fig3}
\end{figure}

The matrix $W$ results  
\begin{eqnarray}
W_{11}&=&\frac{1}{2}(2-p)\,,\quad 
W_{22}= \frac{p}{2}\,.\\
W_{12}&=& 0 =W_{21}^*\,.
\end{eqnarray}
Hence, $S(W)=-(p/2)\log(p/2)-(1-p/2)\log(1-p/2)$.
Some examples of the rate-noise relation are plotted in Fig.~\ref{fig3}.
In particular, Fig.~\ref{fig3}a) refers to the limit case
of pure $\rho_{AB}$ state giving
the concavity of $\chi$ exactly opposite to that of the noise. 
Instead, in Fig.~\ref{fig3}b) one can clearly see an enhancement of $\chi$ for increasing noise (e.g. for $0.15<p<1$). This becomes true for all values of $p$ as soon as the state $\rho_{AB}$ becomes completely mixed. Notice, however that the stochastic resonance effect does not allow the rate to exceed one.

\section{Conclusion}
In conclusion, 
we have shown the possibility of stochastic resonance effects in quantum channels explicitly providing examples where the information transmission rate increases by increasing the noise in the channel.
The first two examples share some analogies with memory effects in quantum channels \cite{mem}
because the stochastic resonance comes out when one error takes place (e.g. bit flip) and another correlated does (e.g. phase flip). This leads to enhancement of coherent information \cite{mem}.

Whether the presented results can be extended to capacities is a subject under investigation, anyway
this work may pave the way to a profitable use of noise in quantum channels.



\begin{references}

\bibitem{GHJM}{See e.g. L. Gammaitoni, P. H\"anggi, P. Jung, and F.Marchesoni, 
{\it Rev.  Mod. Phys.} {\bf 70}, 223 (1998).}

\bibitem{BZ}
A. R. Bulsara and A. Zador, {\it Phys. Rev. E} {\bf 54}, 2158 (1996);
F. Chapeau-Blondeau, {\it Phys. Rev. E} {\bf 55}, 2016 (1997);
J. W. C. Robinson, D. E. Asraf, A. R. Bulsara, 
and M. E. Inchiosa, {\it Phys. Rev. Lett.} {\bf 81}, 2850 (1998).

\bibitem{Keyl}
M. Keyl, Phys. Rep. \textbf{369}, 431 (2002).

\bibitem{Ting}{J. Juhi-Lian Ting, {\it Phys. Rev. E} {\bf 59}, 2801 (1998).}

\bibitem{PLA}{G. A. Bowen and S. Mancini, {\it Phys. Lett. A} {\bf 321}, 1 (2004).}

\bibitem{Kraus}
K. Kraus, \textit{States, Effects and Operations}, (Springer, Berlin, 1983).

\bibitem{Sud}
E. C. G. Sudarshan, P. M. Mathews and J. Rau, Phys. Rev. \textbf{121}, 920 (1961).

\bibitem{MB}{S. Mancini, S. Bonifacio, {\it Phys. Rev. A} {\bf 64}, 032308 (2001).}


\bibitem{garry}
G. Bowen, Phys. Rev. A \textbf{63}, 022302 (2001).

\bibitem{Wer}
R. F. Werner, Phys. Rev. A \textbf{40}, 4277 (1989).

\bibitem{mem}
C. Macchiavello and G. M. Palma, Phys. Rev. \textbf{65}, 050301R (2002);
G. Bowen and S. Mancini, Phys. Rev. \textbf{69}, 012306 (2004).

\end{references}
\end{document}